# Quantification of protein homodimer affinity using native mass spectrometry

Jonathan Schulte ‡ [a], Eric Schwegler ‡ [b], Ute A. Hellmich[b, c, d], Nina Morgner*[a]

[a]    Institute of Physical and Theoretical Chemistry
Goethe-University, Frankfurt
Max-von-Laue-Str. 9, 60438 Frankfurt/Main, Germany

[b]    Faculty of Chemistry and Earth Sciences
Institute of Organic Chemistry and Macromolecular Chemistry,
Friedrich-Schiller-University, Jena, Germany

[c]    Center for Biomolecular Magnetic Resonance (BMRZ)
Goethe-University, Frankfurt/Main, Germany

[d]    Cluster of Excellence Balance of the Microverse
Friedrich-Schiller-University, Jena, Germany

‡shared first authors

*for correspondence: morgner@chemie.uni-frankfurt.de

**ABSTRACT:** Biological processes rely on finely tuned homo- and heteromeric interactions between (biomacro)molecules. The strength of an interaction, typically given by the dissociation constant ($K_D$), plays a crucial role in basic research and must be monitored throughout the development of drugs and agrochemicals. An ideal method for $K_D$ determination is applicable to various analytes with a large range of affinities, tolerates complex matrix compositions, does not require labeling, and simultaneously provides information on the structural integrity of the binding partners. Native mass spectrometry meets these criteria but typically struggles with homooligomeric complexes due to overlapping mass signals. To overcome this, we resolve monomer/dimer contributions to overlapping MS-peaks by separately analyzing the charge state distribution of each oligomeric species via sample dilution and covalent crosslinking. Following this approach, we show that quantitative Laser-Induced Liquid Bead Ion Desorption mass spectrometry (qLILBID-MS) accurately captures the affinities of Bovine Serum Albumin (BSA) and chemically induced dimers of Tryparedoxin (Tpx), an oxidoreductase from human pathogenic *Trypanosoma brucei* parasites, with various molecular glues and homodimer affinities. Conveniently, qLILBID-MS requires a fraction of sample used by other methods such as isothermal titration calorimetry (ITC) and yields previously inaccessible protein homodimer $K_D$s in the high micromolar range, which allowed us to monitor the gradual decrease in homodimer affinity via mutation of crucial dimer interface contacts. Overall, qLILBID-MS is a sensitive, robust, fast, scalable, and cost-effective alternative to quantify protein/protein interactions, that can accelerate contemporary drug discovery workflows, e.g. the efficient screening for proximity inducing molecules like proteolysis targeting chimera (PROTACs) and molecular glues.

## INTRODUCTION

All biological processes including enzymatic catalysis, signal transduction, cellular localization, or the assembly of functional molecular machines, are driven by interactions between (biomacro)molecules across a wide range of affinities.[1–3] The deliberate induction of protein/protein interactions (PPIs) via external stimuli such as light or addition of small molecules is a versatile strategy to exert spatiotemporal control over biological processes.[4,5] Chemically induced dimerization (CID) of proteins has emerged as a particularly powerful strategy to address therapeutic targets in inflammatory, neurodegenerative and infectious diseases that were previously deemed "undruggable".[6,7] Prominent examples are so called molecular glue degraders and proteolysis-targeting chimeras (PROTACs), which initiate the degradation of a biological target via induced proximity to an E3 ubiquitin ligase.[8] The identification of such proximity inducing molecules, and in general the description of biomacromolecular interactions, requires the ability to precisely and efficiently quantify binding affinities, typically through the determination of the corresponding dissociation constants ($K_D$). This is crucial to elucidate structure activity relationships (SARs), to identify and understand key molecular interactions and to provide the basis for drug development.[1,9,10]

Established techniques for protein affinity determination, such as surface plasmon resonance (SPR) or isothermal titration calorimetry (ITC), offer valuable insights but typically involve substantial analyte quantities or elaborate sample preparation.[11–14] This might include an additional step to elucidate biomolecular complex stoichiometries to obtain reliable $K_D$ values.[1,15,16]

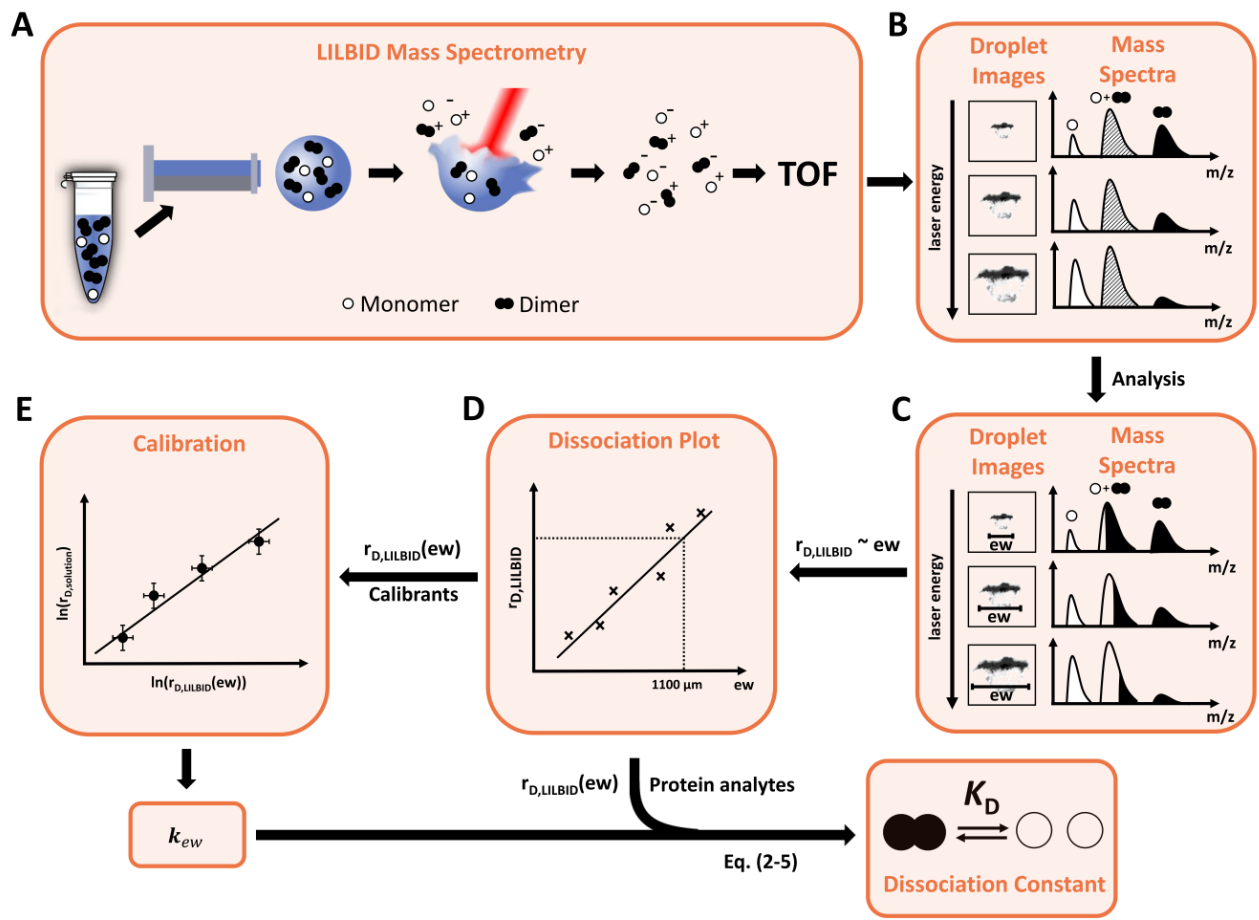

*Figure 1:* Workflow for quantitative LILBID-MS based determination of dissociation constants ($K_D$) of homodimeric biomacromolecules. First, LILBID-MS spectra and corresponding droplet explosion images are recorded at various droplet positions and laser energies. Then, the ratio of dissociation ($r_{D, LILBID}$) is calculated from the MS-peak integrals as the ratio of monomer signal to total signal (see equation 5). Deconvolution of overlapping charge states (e.g. singly charged monomer and doubly charged dimer) is achieved via dilution or crosslinking strategies (detailed in the main text), and droplet explosion widths (ew) are extracted via image analysis. Plotting $r_{D,LILBID}$ against explosion width yields a dissociation plot which is used to deduce $r_{D,LILBID}$ for a certain ew (here: 1100 µm, details in main text) via linear regression. For calibrants with known $K_D$s, this value is used to generate a calibration plot correlating $\ln(r_{D,solution})$ with $\ln(r_{D,LILBID}(ew))$ which yields the calibration factor $k_{ew}$ (see equation 6). This factor is then used to calculate $r_{D,solution}$, and ultimately the dissociation constant $K_D$, from $r_{D,LILBID}(ew)$ of analyte dimers with unknown affinities. Black, white, and hatched peaks in the spectra represent dimers, monomers, and mixed monomer/dimer peaks with overlapping charge states, respectively.

Fluorescence-based methods are versatile and require significantly less material but are dependent on the introduction of fluorophores through non-native tags and/or specific labeling approaches, often enabled by mutagenesis. Finally, tag-free methods such as NMR spectroscopy are less invasive and offer direct insights into analyte integrity but again require isotope labeling and comparatively high amounts of sample. In all cases, homooligomeric systems present particular challenges when aiming to quantify binding affinities, e.g. if the separation of binding partners is required prior to a measurement, as it is usually the case in titration-based methods.

Native mass spectrometry (MS) has become a powerful technique for simultaneously assessing the composition, stoichiometry, and binding affinity of macromolecular complexes without the need for labeling, large sample amounts or extensive sample preparation.[17,18] Its gentle ionization process preserves non-covalent interactions, allowing intact complexes to be detected directly.[17,19] For instance, dissociation constants ($K_D$) can be determined, using nano-electrospray ionization (ESI)-MS titration, where mass spectra are recorded at different analyte concentrations.[20,21] Competitive ligands with known affinities may be added to refine specificity and validate binding interactions.[22,23] Overall, native MS offers several advantages compared to conventional methods,[24,25] including high sensitivity, minimal sample consumption, concurrent determination of identity and integrity of an analyte, simultaneous determination of complex stoichiometry, and the ability to study biomolecular interactions in complex (biological) matrices. [15,26–28]

We recently introduced quantitative laser-induced liquid bead ion desorption (qLILBID)-MS as a novel, native MS-based method for the analysis of stoichiometries and binding affinities of heteromeric biomacromolecular complexes.[28,29] This top-down method enables direct determination of the $K_D$ value from a single sample, eliminating the need for separation of the binding partners and repeated measurements across varying analyte concentrations.[28,29] Here, we extended this approach to homodimeric protein complexes, with $K_D$ values in the low nanomolar to high micromolar range

To establish qLILBID as a versatile tool to investigate diverse biomolecular complexes across a wide range of biologically relevant affinities, several challenges had to be overcome: In LILBID-MS, aqueous sample droplets are generated and transferred into vacuum. Each droplet is irradiated by an IR laser pulse, which excites the OH vibration of water, resulting in explosive droplet expansion during which ionized biomolecular complexes are submitted into the gas phase for mass spectrometric analysis via Time-of-Flight (ToF). (**Figure 1 A**). While most laser energy is absorbed by the sample droplet liquid matrix, some energy induces dissociation of the non-covalent analyte complexes, an effect that becomes more pronounced at higher laser energies.[29,30] This is the effect on which the qLILBID $K_D$ determination is based. The challenge for reproducible measurements is that the energy uptake, and thus the degree of laser induced complex dissociation ($r_{D,LILBID}$), varies from droplet to droplet, due to unavoidable variations in sample droplet trajectories and velocities, thereby affecting the overall laser beam exposure of each droplet. To address this, we correlated the laser energy absorption with the respective droplet plume expansion for each droplet, quantified as the "explosion width" (ew) at 5 µs post-irradiation (**Figure 1 B**).[29] Together, the qLILBID-MS spectra and the droplet explosion widths (ew) then enabled to directly correlate energy transfer and the degree of biomolecular complex dissociation ($r_{D,LILBID}$), the basis for $K_D$ determination (**Figure 1 C**). Taking advantage of the linear relationship between energy uptake ($\propto$ ew) and complex dissociation ($r_{D,LILBID}$), the laser energy is ramped to determine $r_{D,LILBID}$ for a range of droplet explosion widths (**Figure 1 D**), allowing to deduce $K_D$ values by comparing the analyte's $r_D$ $_{LILBID}$ to a calibration plot, generated from $r_{D\ LILBID}$ values of calibration standards recorded under the same experimental conditions (**Figure 1 E**).

Although we previously showed that qLILBID-MS is applicable to determine the affinities of heterodimeric oligonucleotides (double stranded DNA) or RNA/protein complexes[28,29], homodimeric proteins present specific challenges. Here, the low charge state distributions generated in the LILBID process lead to peak overlap in the mass spectra between monomeric ions (M) and dimers (D) with double charge states (same mass/charge ratio (m/z) for $M^{x-}$ and $D^{2x-}$ ions), severely complicating monomer/dimer quantification. With two complementary approaches, i.e. complex dilution and covalent crosslinking using bovine serum albumin (BSA) as a benchmark, we show that MS peak overlap can be resolved

To then put our approach to the test across a wide affinity range, we determined the $K_D$ values of chemically induced homodimers of Tryparedoxin (Tpx), an essential oxidoreductase from human pathogenic *Trypanosoma brucei* (*T. brucei*) parasites.[31] Covalent attachment of different molecular glues to the nucleophilic active site residue cysteine40 induces homodimerization of the protein with low to high micromolar affinities.[32,33] Likewise, mutations in the Tpx dimer interface tune dimer affinity upon binding of the molecular glues[34,35]. This system thereby allowed us to investigate the role of key interactions in the dimer interface using qLILBID.

Here, we establish qLILBID-MS as an accurate tool to capture $K_D$ values of non-covalent protein homodimers, allowing us to sensitively resolve even small variations in $K_D$ values using a comparatively small amount of unlabeled analyte, a key requirement for the reliable determination of biomacromolecular interactions in a physiological affinity range.

## RESULTS AND DISCUSSION

### Correlating peak ratios with solution ratios

The stability of a bimolecular complex in solution is typically given as dissociation constant $K_D$, defined as:

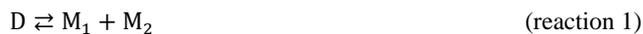

$$K_D = \frac{[M_1] \cdot [M_2]}{[D]} \quad (1)$$

with the concentrations of monomers ($M_1$, $M_2$) and dimers (D) present in a steady-state dissociation equilibrium:

$$D \rightleftarrows M_1 + M_2 \quad \text{(reaction 1)}$$

For homodimers, the definition of the dissociation constant ($K_D$) changes to:

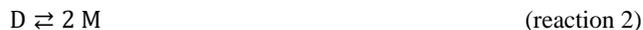

$$K_D = \frac{[M]^2}{[D]} \quad (2)$$

as dimer formation solely depends on the concentration of the monomer (M):

$$D \rightleftarrows 2M \quad \text{(reaction 2)}$$

:

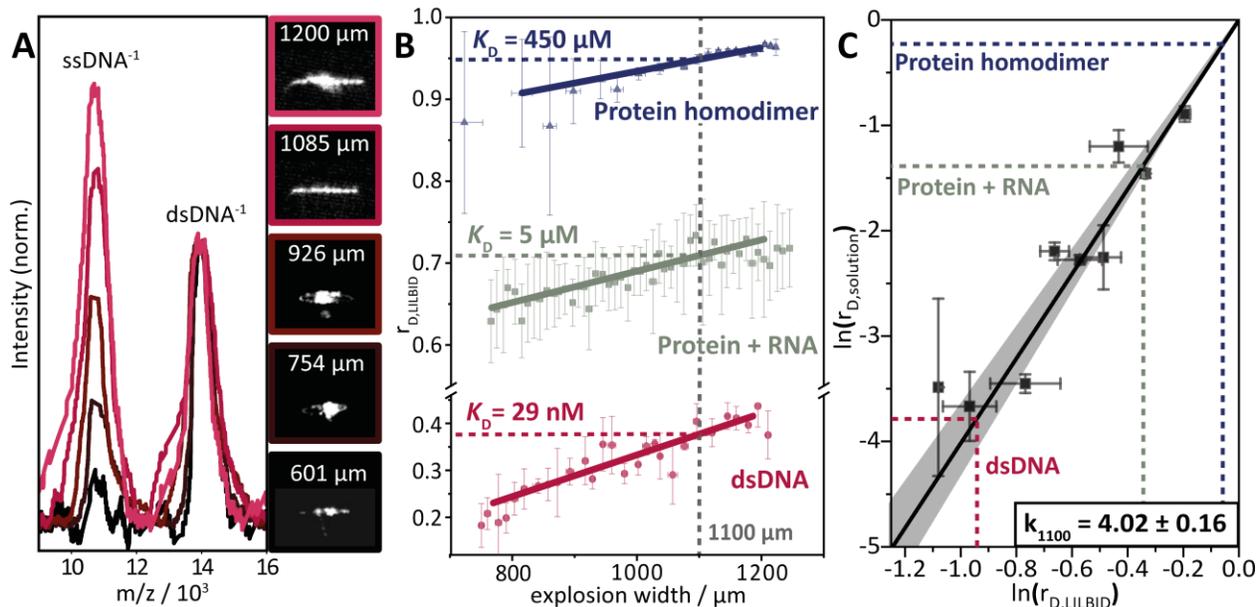

*Figure 2:* Analysis of LILBID-MS spectra and droplet explosions for the $K_D$ determination of diverse biomolecular dimers using a dsDNA calibration standard. (**A**) LILBID-MS spectra (left) and corresponding images of droplet explosions (right, taken 5 µs post IR irradiation) are shown at different laser energy transfer levels. LILBID-MS spectra of dsDNA (strC + cstrC(7-16)[29], see Figure S1) show enhanced dimer dissociation with increasing droplet explosion width, a proxy for the transferred laser energy into the droplet. The measurements were conducted with 10 µM of each DNA strand, and the LILBID-MS spectra were normalized to the height of the dsDNA peak. (**B**) Exemplary dissociation plots of different biomolecular dimers. The dsDNA analyte contains strC + cstrC(7-16) (see **Figure S1**), the "Protein + RNA" sample contains the RNA binding protein Roquin and a linear Roquin binding element (LBE) composed of an 15 nucleotide ssRNA[28], and the "protein homodimer" sample contains the *T. brucei* oxidoreductase Tpx with the covalently bound molecular glue CtFT (see **Figure S4**).[33] The degree of dissociation ($r_{D,LILBID}$) is plotted against the explosion width (ew), and the $r_{D,LILBID}$ at ew = 1100 µm, required for $K_D$ determination, is indicated with dotted lines. (**C**) Calibration plot with the logarithm of $r_{D,solution}$ plotted against the logarithm of $r_{D,LILBID}$. The proportionality factor $k_{1100}$ was obtained via linear regression (black line) according to equation 6,. The calibration points (black) stem from different calibrant dsDNAs with known $K_D$s[29], and have been measured at 10 µM analyte concentration. The grey area represents the confidence interval of 0.95. Based on the experimentally determined $k_{1100}$, the $K_D$ values of DNA, Protein/RNA and protein dimers can be determined by using equations 4a or 4b, and 6.

Due to the dissociation of the dimer into two identical monomers instead of two different compounds, the concentration ratio of monomer to dimer differs for homodimers and heterodimers at a given dissociation constant and analyte concentration ($c_{total}$). To nonetheless establish a consistent workflow for homo- and heterodimers alike, we defined their ratio of dissociation in solution ($r_{D,\,solution}$) as:

$$r_{D,solution} = \frac{n(M_1)}{n(M_1) + n(D)} \quad (3)$$

with $n$ as the number of monomers ($M_1$) and dimers (D), respectively. Together with the total concentration of protein, either free or bound ($c_{total}$), the $r_{D,solution}$ can be used to calculate $K_D$ values of heterodimers using equation 4a

$$K_{D,heterodimer} = \frac{r_{D,\,solution}^2 \cdot c_0}{1 - r_{D,\,solution}} \quad (4a)$$

and $K_D$ values of homodimers using equation 4b (for more details, see equations S10-S16):

$$K_{D,homodimer} = \frac{r_{D,\,solution}^2 \cdot c_0}{r_{D,\,solution}^2 - 3\,r_{D,\,solution} + 2} \quad (4b)$$

As we have previously shown for dsDNA and protein/RNA heterodimers,[28] LILBID-MS can accurately capture the thermodynamic equilibrium state in solution. It has to be noted that the ratio of monomeric versus overall signal in the LILBID spectra does not per se represent the solution ratio (and therewith $r_{D,solution}$), since laser dissociation can increase the number of monomers appearing in the mass spectra. Nevertheless, $r_{D,\,solution}$ can be correlated with the respective ratio in LILBID spectra ($r_{D,\,LILBID}$), defined as:

$$r_{D,LILBID} = \frac{\int M_1 \text{ peaks}}{\int M_1 \text{ peaks} + \int D \text{ peaks}} \quad (5)$$

assuming the ionization efficiency of the different species is comparable. Hereby, "$\int X$ peaks" is the peak integral of species X, here either M1 or D.

For the calculation of peak integrals, the ToF spectrum rather than the m/z spectrum was analyzed, since the ToF spectra reflect the temporal distribution of ion arrivals, effectively counting ions as a function of flight time, while the standardly used m/z spectra replace the ToF axis with m/z with $(m/z)^2 \propto$ ToF. Therefore, the resulting peak integrals of the ToF spectra are better suited to quantify the corresponding species. Further details on the calculation of peak areas for oligonucleotides can be found in Young et al.[16] and Schulte et al.[15]

To deduce the ratio of monomers in solution ($r_{D,solution}$) from the qLILBID spectra, a correction factor had to be defined which correlates $r_{D, solution}$ (taking only solution monomers into account) with $r_{D, LILBID}$ (including solution monomers and monomers which are laser dissociation products) which is directly calculated from the LILBID spectra. To this end, the extent to which a given laser energy transfer level increases the signal stemming from monomeric species had to be determined via calibration. Here, we used a calibration set of dsDNAs with known $K_D$ values (see below) and determined the correction factor ($k_{ew}$) for the given energy transfer.

Thus, based on $r_{D,LILBID}$ at a given explosion width (ew), $r_{D,solution}$ (equation 3) can be calculated using equation 6:

$$\ln(r_{D,solution}) = \ln(r_{D,LILBID}) \cdot k_{ew} \tag{6}$$

with the corresponding proportionality factor $k_{ew}$. This allowed us to consider dissociation during the LILBID process, and determine $r_{D, solution}$ and thus the $K_D$ of a hetero- or homodimeric analyte using equation 4a or 4b.

**Calibration**

For calibration, we used a set of previously reported[29] dsDNA heterodimers with nM to low μM affinities t containing three different DNA strands with a length of 35 nucleobases (strA, strB, strC) and 9 complementary, shorter strands with 8 to 15 nucleobases (cstrX, **Figure S1**).

For all heterodimeric dsDNAs, LILBID-MS spectra were measured at varying laser intensities and for each spectrum, the proportions of dissociated DNA ($r_{D, LILBID}$) were determined. Each data set contained the $r_{D,LILBID}$ value and the correlated droplet explosion width (ew) 5 μs after irradiation (**Figure 2A**). Plotting $r_{D, LILBID}$ against ew yielded dissociation plots similar to those shown in **Figure 2B** (red plot). For all dsDNAs, we found a linear relationship between $r_{D, LILBID}$ and the ew between 820-1200 μm (**Figure S2**). In this ew range, the laser desorption was strong enough to produce a sufficient ion count but no ions were lost due to the geometry of the ion optics.

For precise calibration, the dissociation behavior of all calibrant dsDNAs had to be assessed at the same explosion width (ew) to ensure that a comparable laser energy transfer occurred in all analyte droplets. Since a large laser energy transfer favors high ion counts, a high explosion width in the linear region should be selected to obtain mass spectra with high reliability. Here, we chose an ew of 1100 μm and determined the $r_{D,LILBID}$ values for all calibrant dsDNAs from their respective, linearly fitted dissociation plots (**Figure S2**). **Figure 2C** shows the calibration plot, with the logarithm of these $r_{D,LILBID}$ values plotted against the logarithm of the orresponding $r_{D,solution}$ values, which were calculated with $K_D$s from literature using equation 4a.[29] Finally, for the given set of calibrants and experimental setup, a linear fit yielded the proportionality factor $k_{1100}= 4.02 \pm 0.16$ which was used for all following $K_D$ determinations.

**$K_D$ values for homodimers - Method validation**

In all previous studies that employed qLILBID-MS for $K_D$ determination, the biomolecular analytes formed heterodimers and had a large mass difference between binding partners. This prevents an overlap in charge states[28,29,36], allowing for an unambiguous assignment of peak integrals to the different analyte species. In addition, binding a much smaller interaction partner barely affects the ionization efficiency, and thus the charge state distribution, of the larger monomer which allowed to obtain $r_{D, LILBID}$ from the integrals of singly charged MS peaks from the larger monomer and the heterodimeric complex. However, homomeric interactions are very common in biomacromolecules, thus presenting the urgent need to extend the use of qLLIBID-MS also to such systems.

To adapt the qLILBID method for homodimers, we had to consider all charge states of monomer and dimer species, as well as charge state overlaps of e.g. singly charged monomer and doubly charged dimer with identical m/z. To account for the latter, we pursued two different strategies, crosslinking and dilution, i. e. to either create permanent dimers or strongly suppress dimerization. Using the well-characterized model protein bovine serum albumin (BSA) with a homodimer $K_D$ of $10 \pm 2$ μM as a standard,[37] we could determine the charge state distributions of both dimer and monomer. Knowing the charge state ratio of either the monomeric or the dimeric species then sufficed to calculate their relative abundance in the overlapping MS peaks.

**Crosslinking approach**

In LILBID-MS spectra recorded with BSA solution at a concentration of 30μM (**Figure 3A**, red spectrum), we mainly observed monomers with one to four negative charges, but also smaller peaks that indicated BSA dimers ($BSA_2^{1-}$, $BSA_2^{3-}$). From this, we inferred that additional dimer peaks ($BSA_2^{2-}$, $BSA_2^{4-}$) overlapped with peaks of the BSA monomer ($BSA^{1-}$, $BSA^{2-}$). To quantify the relative abundances of each species within the overlapping peaks, we analyzed the charge state distribution of isolated dimer signals by crosslinking BSA with the reagent 1-ethyl-3-(3-dimethylaminopropyl)carbodiimide (EDC) and removing monomeric protein via filtration prior to the qLILBID-MS analysis. (**Figure 3A**, bottom). Next, we determined the relevant peak areas for these charge states (-1 to -4) and scaled this pattern to the "dimer only" peaks observed in the MS spectrum of unmodified BSA ($BSA_2^{-1}$ and $BSA_2^{3-}$ at m/z 132k and 44k respectively). Subtracting the scaled BSA dimer contribution revealed the peak intensities attributable to monomers. As laser energy transfer influences the analyte's ionization, and thus the charge state distribution of each species, we determined a charge state correlation function $cs_{corr}(ew)$. It describes the linear charge state distribution shift in dependence of ew (**Figure S3**). Importantly, MS spectra of crosslinked and unmodified BSA have to be compared for the same ew (see Supporting Information for details (equations S1-9 and **Figure S3A-B**).

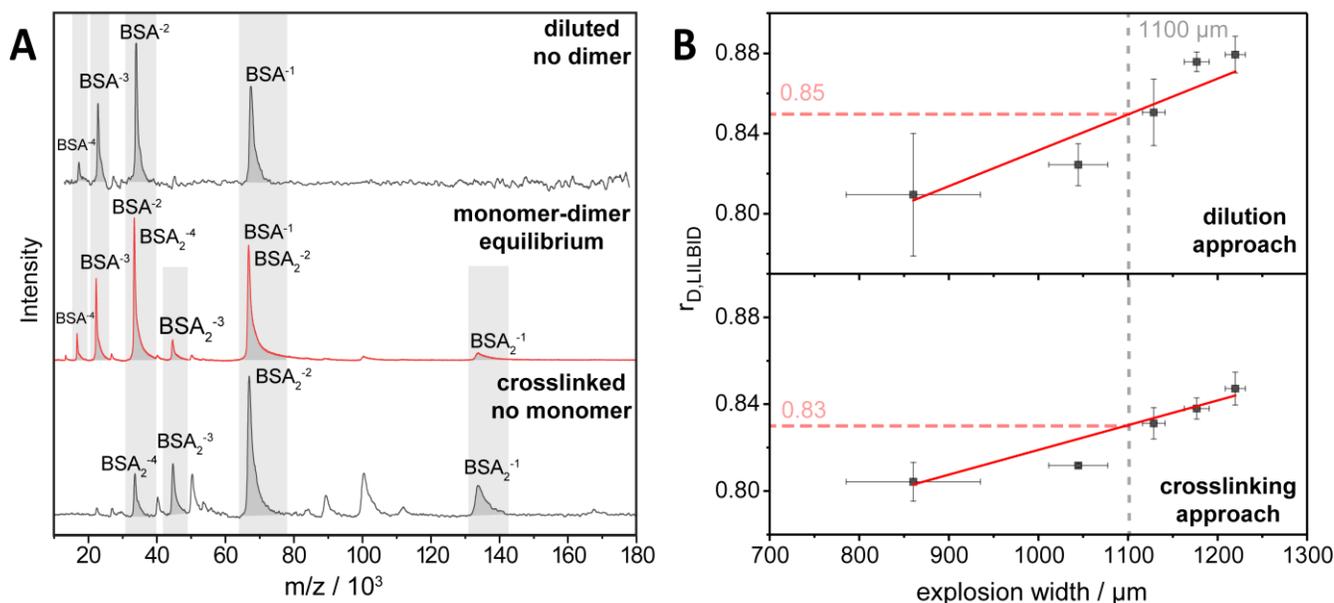

*Figure 3:* Crosslinking and dilution approaches can be used to disentangle overlapping charge states of protein monomers and dimers. (**A**) MS-spectra of BSA, diluted to 100 nM (top, black), 30 µM BSA (middle, red) and 30 µM crosslinked BSA (bottom, black). MS-peaks corresponding to BSA monomers and dimers are highlighted in grey. Either the spectra of diluted BSA or of crosslinked BSA dimers can be used to determine the amount of monomer and dimer in the MS spectra of unmodified BSA, and calculate $r_{D,LILBID}$ with equation 5. Additional peaks in the crosslinked spectra stem from higher oligomers and do not interfere with the analysis. (**B**) BSA dissociation plots obtained following the dilution (top) or the crosslinking approach (bottom) to disentangle overlapping MS peaks in the same MS spectrum of 30 µM BSA (**A**, red). $r_{D,LILBID}$ was plotted against the explosion width and linearly fitted between 800 µm and 1250 µm to obtain $r_{D,LILBID}$ at an ew of 1100 µm for $K_D$ determination with equation 4b and 6 (for details, see main text).

We determined the $r_{D,LILBID}$ for BSA at different explosion widths between 750 and 1250 µm, and interpolated an $r_{D,LILBID}$ of 0.83 ± 0.03 for ew= 1100 µm (**Figure 3B**, bottom). With the $k_{1100}$ value established with the dsDNA calibrants and equation 6, we calculated a $r_{D,solution}$ of 0.42 ± 0.07. Using equation 4b and $c_{total}$= 30 µM, we ultimately obtained a BSA dimer $K_D$ of 9 ± 3 µM which is in good agreement with the literature value of 10 ± 2 µM.[37]

**Dilution approach**

To bias the sample composition to the monomeric state, we analyzed a diluted BSA solution with a concentration of 100 nM. This is well below the reported BSA dimer $K_D$, and yields less than 1 % dimer in solution, thereby allowing the unambigous and determination of the the charge state distribution of BSA monomers (**Figure 3A**, top,). Following this approach, we were able to reliably disentangle the spectral contributions of BSA monomers and dimers at the higher protein concentration (30 µM, **Figure 3A**, red spectrum). As for the crosslinking approach, we considered the effect of the laser energy transfer on the analyte's ionization, and recorded MS spectra of 100 nM BSA at explosion widths between 600 and 1200 µm (**Figure S3A**). By scaling the charge state distribution of the purely monomeric species based on non-overlapping peaks, similar as previously as done wth the crosslinked dimers, we retrieved the separated signal intensities stemming from monomers and dimers, respectively. This yielded $r_{D,LILBID}$ values which were linearly fitted (**Figure 3B**, top) to obtain an $r_{D,LILBID}$ of 0.85 ± 0.02 at an ew of 1100 µm. Using equation 6 and $k_{1100}$ for our experimental setup, we calculated an $r_{D,solution}$ of 0.54 ± 0.21, which, together with equation 4b and $c_0$= 30 µM, yielded a $K_D$ of 11 ± 4 µM. Again, our approach was in good agreement with the literature value of 10 ± 2 µM.[37]

In summary, both the crosslinking and dilution approach yielded $K_D$ values (9 ± 3 µM and 11 ± 4 µM, respectively) matching the reported literature value (10 ± 2 µM) for BSA dimerization.[37] Our results thus show that both strategies can be used to reliably quantify homodimer affinities with LILBID-MS. Overall, the dilution approach was more sample and time efficient than crosslinking, as it did not require prior chemical modification. However, the dilution approach is in general only applicable to samples with sufficiently low binding affinities since complex dissociation has to occur at sample concentrations that do not undercut the LILBID detection limit.[30] Further, dilution decreases the signal to noise ratio and thus increases the method's intrinsic error. Therefore, the crosslinking approach is preferable when a sufficient amount of sample is available.

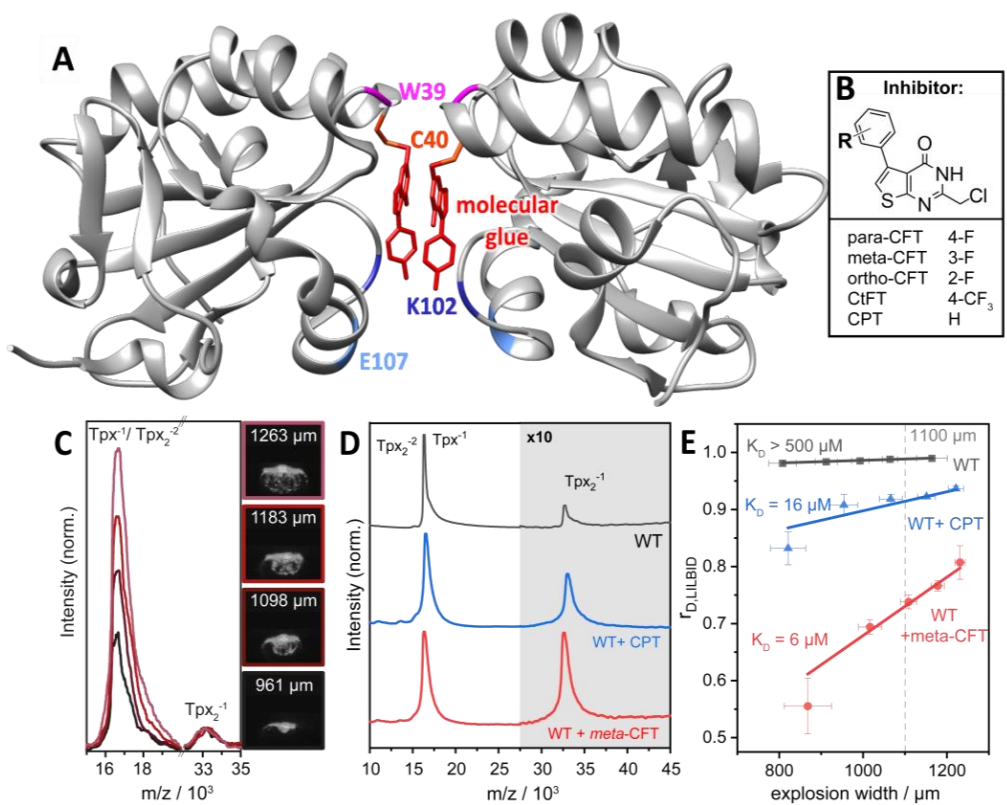

*Figure 4:* qLILBID-MS accurately captures affinities of chemically induced Tpx homodimers with different molecular glues. (**A**) X-ray crystal structure of the chemically induced Tpx dimer (PDB: 6GXG, chains A, B).[32] The dimer interface consists of a molecular glue (here 2-(chloromethyl)-5-(4-fluorophenyl)thieno[2,3-*d*]pyrimidin-4(3*H*)-one, "para-CFT") covalently bound to Tpx residue C40 (red), two tryptophane residues (W39, pink), and salt bridges between the charged residues (K102 and E107, dark and light blue, respectively) (all sidechains shown in **Figure 5**). The tertiary structure of the protein backbone is shown in grey. (**B**) Structural formulas of multiple molecular glues that induce Tpx dimerization upon covalent binding to cysteine residue 40.[33] (**C**) LILBID-MS spectra of Tpx WT with covalently bound meta-CFT (left) and enhanced images of droplet explosions with varying sizes taken 5 µs post IR irradiation (right). The measurements were conducted at a concentration of 30 µM Tpx and the spectra were normalized to the peak height of the dimer peak at m/z= 33 · 10³. (**D**) LILBID-MS spectra of Tpx WT without inhibitor (top, grey), or with the molecular glues CPT (blue, middle), and meta-CFT (red, bottom). All graphs are the sum of 98 spectra, each resulting from one droplet's explosive expansion. The sample concentration was 30 µM and the area around the dimer peak is displayed with a tenfold intensity (grey background). (**E**) Exemplary dissociation plots of unmodified Tpx WT (grey, top), Tpx WT +CPT (blue, middle), and Tpx WT +meta-CFT (red, bottom). The dotted line marks the explosion width of 1100 µm which is used to extrapolate $r_{D,LILBID}$ for the $K_D$ determination. The annotated $K_D$ is the result of three measurements (**Figure S4**) and was calculated using equations 4b and 6.

**Determination of chemically induced protein homodimer affinities using qLILBID-MS**

Encouraged by the accurate determination of BSA homodimer affinity, we next sought to benchmark qLILBID-MS using a set of chemically inducible protein homodimers. As we previously reported, the homodimerization of the *T. brucei* oxidoreductase Tpx is strictly contingent on the interaction with small molecular glues derived from the covalent inhibitor para-CFT (**Figure 4A, B**).[32,33] Binding of these molecular glues to the active site cysteine40 results in the formation of Tpx homodimers with $K_D$ values in the micromolar range.[32,33] In addition to small molecule contacts (**Figure 4A, 5D**), the Tpx homodimer interface consists of two tryptophane sidechains (W39, **Figure 4A, 5D**) and two salt bridges (between K102 and E107 from another chain, **Figure 4A, 5A**). Point mutations of either of these residues in the dimer interface, and changes in the fluorination pattern of the molecular glue, can tune Tpx homodimer affinity by two orders of magnitude in the micromolar range[32,33,35] (**Table 1**). Hence, the Tpx-based CID system constitutes an excellent standard to benchmark our qLILBID-MS method.

Due to the sufficient availability of purified protein, we first decided to pursue the crosslinking approach to determine the charge state distribution of crosslinked Tpx homodimers at different explosion widths (**Figure S3C**). Using the resulting charge state correlation function from the measurement of crosslinked Tpx allowed us to extract the $r_{D,LILBID}$ from MS spectra of Tpx samples with mixed monomer/dimer populations.

Next, we analyzed the effect of different molecular glues (para-CFT, meta-CFT, CPT, ortho-CFT, and CtFT, see **Figure 4B**) on the formation of chemically induced homodimers of Tpx wild type (WT). To this end, we followed the previously established qLILBID workflow (**Figure 1, 4C-D**), fitted the resulting dissociation plots (**Figure 4E and S4**), determined $r_{D,LILBID}$ at an ew of 1100 µm, and calculated the Tpx homodimer affinities ($K_{D,LILBID}$) using equations 4b and 6 (**Figure 4E, Table 1, Supplementary Table 2, 3**).

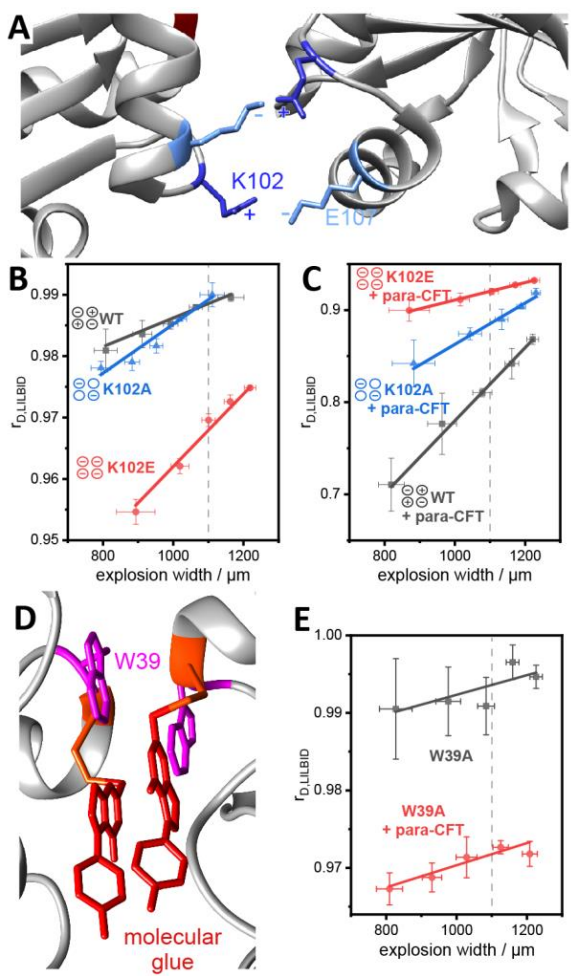

*Figure 5:* (**A**) Zoom into the Tpx dimer interface highlighting sidechains of Tpx residues K102 and E107, which form intermolecular salt bridges that stabilize the chemically induced Tpx dimer interface.[32] (**B**) Dissociation plots of Tpx WT, K102A and K102E. Without inhibitor, all constructs are predominantly monomeric. (**C**) Dissociation plots of para-CFT bound Tpx WT, K102A and K102E. As anticipated, increasing the negative charge in the dimer interface leads to a decrease in chemically induced homodimer affinity (WT > K102A > K102E). To determine the respective $K_D$ values, $r_{D,LILBID}$ was determined at an ew of 1100 µm for all species (dotted lines). (**D**) Zoom into the Tpx dimer interface highlighting sidechains of Tpx residues W39 from both protomers (pink), which contribute to hydrophobic interactions in the dimer interface. (**E**) Dissociation plots of Tpx W39A and W39A+para-CFT. As expected, removing the indole sidechains from the interface greatly reduced homodimer affinity.

Without the molecular glue, Tpx WT shows a very limited propensity to dimerize which manifested in very low or absent dimeric signals and an apparent $K_{D,Lilbid}$ >500µM, i.e. negligible dimerization. The various molecular glue molecules produced a distinct set of $K_D$ values, between $5 \pm 4$ µM (para-CFT) and $370 \pm 110$ µM (CtFT), clearly reflecting their varying affinities. Comparison between the determined affinities by qLILIBID, ($K_{D,LILBID}$) and dilution-ITC ($K_{D,ITC}$) shows that qLILBID-MS accurately captured Tpx homodimer affinities (**Table 1**), requiring significantly less time and sample material, thus establishing qLILBID as a powerful tool for the analysis of non-covalent protein homodimers over a physiologically relevant affinity range. This prompted us to speculate whether the method would also be suited to readily screen the effect of mutations in a dimer interface.

To this end, we mutated residues W39 and K102 to alanine, and K102 additionally to glutamate to create charge repulsion. Upon adding para-CFT, all Tpx mutants were analyzed with qLILBID-MS (**Figure 5**, **Table 1**).

The Tpx W39A mutation was introduced to disrupt key hydrophobic interactions in the dimer interface, as gauged by quantitative studies using SEC, SEC-MALS, SEC-SAXS and $^{19}$F-NMR.[32–35] Indeed, the $K_{D,LILBID}$ value forTpx W39A shows the mutation to causes significantly lower dimer affinity compared to the wildtype protein. Since ITC requires sample concentrations well above the $K_D$.attempts to determine the Tpx W39A dimer affinity with dilution ITC method failed l.[32] In contrast, qLILBID-MS provided a $K_D$ value of $300 \pm 110$ µM for these low-affinity homodimers (**Table 1, Figure 5D, E, Figure S5**), and in good agreement with estimates from other methods.[32–35]

Next, we investigated the role of ionic interactions between lysine residues 102 and glutamate residues 107 from each Tpx protomer (**Figure 5A**). To decrease dimer affinity through the stepwise removal of charge complementarity, we mutated residue K102 either to alanine (K102A) to remove ionic interactions, or to glutamate (K102E) to create repulsive interactions in the dimer interface. For non-molecular glue bound Tpx mutants, the affinities were in the hundreds of µM concurrent with no efficient dimer formation in the absence of a molecular glue. Upon molecular glue binding the gradual decrease of charge complementarity between WT, K102A and K102E mutants was reflected in the resulting decrease in homodimer affinities determined with qLILBID-MS. While Tpx WT in complex with para-CFT had $K_D$ of $5 \pm 4$ µM, the K102 alanine or glutamate mutants decreased the affinity to $K_D$s of $37 \pm 21$ µM and $71 \pm 16$ µM, respectively observed for the Tpx WT with intact ionic interactions (**Figure 5 B-C, Figure S5, Table 1**).

*Table 1:* $K_D$ values of Tpx homodimers obtained from qLILBID-MS and previous ITC measurements.

| Tpx | molecular glue | $K_{D,LILBID}$ [µM] | $K_{D,ITC}$ [µM] [32,33] |
|---|---|---|---|
| **WT** | none | >500 | - |
| | para-CFT | $5 \pm 4$ | $5.3 \pm 1.9$ |
| | meta-CFT | $6 \pm 5$ | $3.7 \pm 1.5$ |
| | CPT | $16 \pm 10$ | $6.0 \pm 2.2$ |
| | ortho-CFT | $26 \pm 13$ | $32 \pm 7$ |
| | CtFT | $370 \pm 110$ | $410 \pm 56$ |
| **W39A** | none | >500 | - |
| | para-CFT | $300 \pm 110$ | - |
| **K102A** | none | >500 | - |
| | para-CFT | $37 \pm 21$ | - |
| **K102E** | none | $310 \pm 130$ | - |
| | para-CFT | $71 \pm 16$ | $83 \pm 32$ |

In summary, we could use qLILBID-MS to determine the dimer $K_D$s of a diverse set of protein homodimers that vary in their structural composition and dimer affinity in a fast and sample-efficient manner.

## CONCLUSION

This study demonstrates that qLILBID-MS is an excellent method to determine (induced) protein dimer affinities in physiologically relevant ranges and therefore complements other available methods. Furthermore, we showed that in some cases, such as chemically induced Tpx dimer, qLILBID can provide results where other methods, such as ITC, fail due to technical limitations. As homodimers present a particular challenge for $K_D$ analysis with qLILBID-MS, we developed two complementary strategies, protein crosslinking and sample dilution, to resolve overlapping MS peaks arising from multiple oligomeric species with identical m/z values. Analyzing BSA and chemically induced Tpx homodimers, we found that both approaches yield $K_D$ values that are in good agreement with the literature.[32,33,35] This set of validation methods should allow researchers to assess the affinities of their biomacromolecules of interest in a straight forward manner, even when faced with limited sample amounts or low affinitiy homodimers.

To test whether qLILBID-MS is a robust method that can detect small variations in dimer affinity, we generated a set of chemically induced Tpx homodimers with various molecular glues and affinities for systematic benchmarking. Small chemical modifications of the inhibitor resulted in shifts in the induced $K_D$ value. Following the herein established protocol, we show that qLILBID-MS accurately reflects our previously reported dimer $K_D$s[33] for the given set of molecular glues over two orders of magnitude.

Tpx dimerization similarly hinges on the interplay of amino acid sidechains in the dimer interface, i.e. W39, K102 and E107 Importantly, qLILBID was able to determine low Kd values inaccessible to ITC..[32] In line with the expected structural contributions of the W39 in the Tpx interface, a $K_D$ shift of almost 2 orders of magnitude compared to Tpx WT was observed. This is in good agreement with estimates from prior studies that employed $^{19}$F-NMR spectroscopy.[33,35] Our results also reinforce the importance of ionic interactions in the Tpx dimer interface, particularly between residues K102 and E107.[32] By mutating K102 to alanine and glutamate, we progressively disrupted charge complementarity and assessed the impact on homodimer affinity via qLILBID-MS. The K102A mutation eliminated energetically favoured salt bridges, which is reflected in a 5-fold decrease in homodimer affinity. Charge inversion achieved with the K102E mutation further destabilized the interface through electrostatic repulsion, causing a 10-fold decrease in affinity compared to Tpx wildtype.

In addition to its accuracy and robustness, qLILBID-MS offers rapid analysis of $K_D$ values for both homo- and heterodimeric biomolecules, while requiring minimal sample consumption. Typical $K_D$ determination by qLILBID requires less than 700 pmol of sample, which inclues measurements in triplicates and all calibration standards (crosslinking or dilution approach.

Importantly, the protein complexes investigated herein exhibited affinities in the micromolar range, which is characteristic of many physiologically relevant protein–protein interactions.[1,38] Notably, this includes high $K_D$ values indicative of weak affinities in the low µM range. While such interactions are often biologically meaningful, e.g. in transient signalling interactions, they are notoriously difficult to quantify using conventional methods. Our approach can distinguish between small differences in $K_D$, driven e.g. by the loss of a salt bridge, or the small modification of a molecular glue, an important prerequisite for dissecting biological interactions or drug development. Together with our previous studies on DNA and protein/RNA heterodimers,[28,29] this highlights the broad applicability of qLILBID-MS that allows a fast and precise determination of $K_D$ values for a plethora of biomacromolecular complexes.

Future improvements in mass resolution, along with continued methodological advancements, will aim to enable the determination of $K_D$ values in increasingly complex systems including heterogeneous samples, higher-order oligomeric assemblies, integral membrane proteins, and lipid/protein interactions or determine the influence of small modifications as post translational modifications (PTMs) on affinities. Ideally, these advances will also support high-throughput capabilities, thereby broadening the method's applicability across diverse areas of structural and functional biology.

## ASSOCIATED CONTENT

Supporting Information.pdf

## AUTHOR INFORMATION

Corresponding Author
* morgner@chemie.uni-frankfurt.de.

Author Contributions
The manuscript was written through contributions of all authors. / All authors have given approval to the final version of the manuscript. / ‡These authors contributed equally.

## ACKNOWLEDGMENT

We acknowledge Dr. Marco D. Preuss and Dr Philipp Klein for the synthesis of molecular glues.
Supported by the Deutsche Forschungsgemeinschaft (DFG, German Research Foundation) via Research training group (RTG) 1986 CLiC (fellowship, to JS) and RTG2015 (fellowship to ES), individual grants (Project-ID 426191805 to NM and 438511573 to UAH), the Collaborative research center (CRC)1507 (Project ID *450648163,* to NM and UAH) and the Cluster of Excellence 'Balance of the Microverse' (Project-ID 390713860, to UAH).

*Supporting Information*

# Affinity of chemically induced protein homodimers quantified by LILBID-MS

Jonathan Schulte[‡] [a], Eric Schwegler[‡] [b], Ute A. Hellmich[b, c, d], Nina Morgner*[a]

| | |
|---|---|
| [a] | Institute of Physical and Theoretical Chemistry |
| | Goethe-University, Frankfurt/Main, Germany |
| [b] | Faculty of Chemistry and Earth Sciences |
| | Institute of Organic Chemistry and Macromolecular Chemistry, |
| | Friedrich-Schiller-University, Jena, Germany |
| [c] | Center for Biomolecular Magnetic Resonance (BMRZ) |
| | Goethe-University, Frankfurt/Main, Germany |
| [d] | Cluster of Excellence Balance of the Microverse |
| | Friedrich-Schiller-University, Jena, Germany |

[‡]shared first authors

*for correspondence: morgner@chemie.uni-frankfurt.de

**Table of Contents**

**Table S1: Linear fits from dissociation plots for the affinity calibration.**

Fitting parameters obtained from the qLILBID calibration using the calibration set (**Figure S1**). For each set of binding partners the monomer-to-complex ratio ($r_{D,LILBID}$, equation (S9)) was extracted from the mass spectra across a range of explosion width and a linear fit was applied (**Figure S2**). From these fits $r_{D,LILBID}$ was determined for an explosion width of 1100 μm. Available $K_D$s from the literature[1] were used to calculate $r_{D,solution}$ values (equation (S3)). Plotting $r_{D,solution}$ against $r_{D,LILBID}$ yields a calibration plot (**Figure 2C**), which was used to determine $K_D$ values throughout this study.

| binding partners | fit - slope | fit - y-intercept | $r_{D,LILBID}$ @ 1100 μm | $K_{D,ITC}$ / nM[1] | $r_{D,solution}$ |
|---|---|---|---|---|---|
| **strA + cstrA(16-26)** | 7.00E-04 ± 6.72E-05<br>8.47E-04 ± 1.78E-04<br>5.58E-04 ± 1.47E-04 | -0.168 ± 0.064<br>-0.408 ± 0.175<br>0.104 ± 0.142 | **0.615 ± 0.080** | 370 ± 50 | 0.45 ± 0.02 |
| **strA + cstrA(18-26)** | 2.26E-04 ± 2.73E-05<br>1.31E-04 ± 7.15E-05<br>2.52E-04 ± 8.43E-05 | 0.580 ± 0.023<br>0.577 ± 0.065<br>0.642 ± 0.081 | **0.823 ± 0.081** | 8500 ± 700 | 0.904 ± 0.007 |
| **strA + cstrA(21-35)** | 3.53E-04 ± 7.06E-05<br>4.35E-04 ± 6.24E-05<br>4.93E-04 ± 4.84E-05 | 0.051 ± 0.069<br>0.089 ± 0.057<br>-0.156 ± 0.049 | **0.464 ± 0.076** | 31.2 ± 0.8 | 0.162 ± 0.002 |
| **strA + cstrA(26-35)** | 6.25E-04 ± 3.43E-05<br>4.14E-04 ± 1.68E-04<br>4.82E-04 ± 1.10E-04 | 0.160 ± 0.033<br>0.011 ± 0.158<br>0.103 ± 0.105 | **0.649 ± 0.157** | 3900 ± 500 | 0.83 ± 0.02 |
| **strB + cstrB(5-16)** | 5.20E-04 ± 9.22E-05<br>6.58E-04 ± 6.83E-05<br>9.07E-04 ± 9.64E-05 | -0.164 ± 0.083<br>-0.397 ± 0.065<br>-0.593 ± 0.086 | **0.380 ± 0.038** | 20.1 ± 1.8 | 0.132 ± 0.005 |
| **strB + cstrB(6-15)** | 3.81E-04 ± 6.44E-05<br>4.88E-04 ± 1.02E-04<br>4.38E-04 ± 6.40E-05 | 0.085 ± 0.060<br>0.139 ± 0.096<br>0.032 ± 0.060 | **0.564 ± 0.079** | 353 ± 7 | 0.443 ± 0.003 |
| **strB + cstrB(8-16)** | 5.34E-04 ± 1.01E-04<br>5.74E-04 ± 9.43E-05<br>7.18E-04 ± 1.45E-04 | 0.114 ± 0.088<br>0.074 ± 0.083<br>-0.051 ± 0.122 | **0.715 ± 0.017** | 2120 ± 60 | 0.741 ± 0.004 |
| **strC + cstrC(7-16)** | 4.33E-04 ± 4.16E-05<br>3.95E-04 ± 6.17E-05<br>4.02E-04 ± 6.06E-05 | -0.119 ± 0.039<br>-0.133 ± 0.061<br>-0.082 ± 0.059 | **0.339 ± 0.027** | 29 ± 7 | 0.16 ± 0.02 |
| **strC + cstrC(8-15)** | 2.07E-04 ± 3.96E-05<br>1.87E-04 ± 7.76E-05<br>5.11E-04 ± 9.69E-05 | 0.239 ± 0.035<br>0.308 ± 0.072<br>0.003 ± 0.086 | **0.516 ± 0.040** | 418 ± 16 | **0.470 ± 0.006** |

**Table S2: Linear fits from dissociation plots of Tpx wild type with and without different molecular glues.**

Fitting parameters obtained from the qLILBID measurements. For Tpx WT (with or without molecular glue) the monomer-to-complex ratio ($r_{D,LILBID}$, equation (S9)) was extracted from the mass spectra across a range of explosion width and a linear fit was applied (**Figure S4**). From these fits $r_{D,LILBID}$ was determined for an explosion width of 1100 µm. The fit from calibration plot (**Figure 2C**) allowed to determine $r_{D,\text{solution}}$ with equation (6) and therewith $K_{D,LILBID}$ with equation (S16c). $K_D$ Values in the range for which we had reference molecules with known affinities from ITC measurements are printed in black, those values outside of the ITC-validated range are printed in grey.

| Tpx WT + molecular glue | fit - slope | fit - y-intercept | $r_{D,LILBID}$ @ 1100 µm | $r_{D,\text{solution}}$ | $K_{D,LILBID}$ / µM |
|---|---|---|---|---|---|
| **WT** | 2.01E-05 ± 6.78E-06 | 0.969 ± 0.007 | 0.992 ± 0.003 | 0.967 ± 0.010 | 810 ± 280 |
|  | 2.34E-05 ± 4.28E-06 | 0.963 ± 0.005 |  |  |  |
|  | 1.27E-05 ± 2.12E-06 | 0.981 ± 0.002 |  |  |  |
| **WT + para-CFT** | 9.66E-04 ± 1.55E-04 | -0.332 ± 0.186 | 0.798 ± 0.048 | 0.404 ± 0.098 | 5 ± 4 |
|  | 4.54E-04 ± 1.51E-05 | 0.344 ± 0.017 |  |  |  |
|  | 3.97E-04 ± 9.83E-06 | 0.383 ± 0.011 |  |  |  |
| **WT + meta-CFT** | 1.15E-04 ± 4.58E-05 | 0.742 ± 0.053 | 0.809 ± 0.058 | 0.426 ± 0.123 | 6 ± 5 |
|  | 4.49E-04 ± 1.38E-04 | 0.333 ± 0.151 |  |  |  |
|  | 5.12E-04 ± 6.67E-05 | 0.167 ± 0.075 |  |  |  |
| **WT + ortho-CFT** | 4.38E-04 ± 4.04E-05 | 0.381 ± 0.047 | 0.897 ± 0.024 | 0.647 ± 0.070 | 26 ± 13 |
|  | -1.43E-05 ± 4.22E-05 | 0.927 ± 0.051 |  |  |  |
|  | 3.32E-04 ± 1.48E-04 | 0.552 ± 0.171 |  |  |  |
| **WT + CPT** | 1.67E-04 ± 4.76E-05 | 0.731 ± 0.055 | 0.871 ± 0.033 | 0.574 ± 0.087 | 16 ± 10 |
|  | 4.95E-04 ± 8.56E-05 | 0.290 ± 0.098 |  |  |  |
|  | 3.75E-05 ± 7.78E-05 | 0.822 ± 0.078 |  |  |  |
| **WT + CtFT** | 5.20E-05 ± 7.14E-06 | 0.922 ± 0.008 | 0.983 ± 0.004 | 0.934 ± 0.016 | 370 ± 110 |
|  | 7.14E-05 ± 4.93E-06 | 0.910 ± 0.005 |  |  |  |
|  | 5.13E-05 ± 1.69E-05 | 0.924 ± 0.019 |  |  |  |

**Table S3: Linear fits from dissociation plots of Tpx mutants with and withot para-CFT**

Fitting parameters obtained from the qLILBID measurements. For each Tpx variant (with or without molecular glue) the monomer-to-complex ratio ($r_{D,LILBID}$, equation S9) was extracted from the mass spectra across a range of explosion width and a linear fit was applied (**Figure S5**). From these fits $r_{D,LILBID}$ was determined for an explosion width of 1100 µm. The fit from calibration plot (**Figure 2C**) allowed to determine $r_{D,solution}$ with equation (6) and therewith $K_{D,LILBID}$ with equation (S16c). $K_D$ Values in the range for which we had reference molecules with known affinities from ITC measurements are printed in black, those values outside of the ITC-validated range are printed in grey..

| Tpx variant + molecular glue | fit - slope | fit - y-intercept | $r_{D,LILBID}$ @ 1100 µm | $r_{D,solution}$ | $K_{D,LILBID}$ / µM |
|---|---|---|---|---|---|
| **K102A** | 3.99E-05 ± 5.31E-06 | 0.945 ± 0.005 | 0.990 ± 0.002 | 0.962 ± 0.007 | 700 ± 140 |
|  | 3.92E-06 ± 2.26E-06 | 0.988 ± 0.003 |  |  |  |
|  | 3.70E-05 ± 3.84E-06 | 0.948 ± 0.005 |  |  |  |
| **K102A + para-CFT** | 1.41E-04 ± 1.67E-05 | 0.751 ± 0.019 | 0.914 ± 0.027 | 0.697 ± 0.083 | **37 ± 21** |
|  | 9.40E-05 ± 6.35E-06 | 0.847 ± 0.007 |  |  |  |
|  | 2.24E-04 ± 2.33E-05 | 0.639 ± 0.027 |  |  |  |
| **K102E** | 1.90E-05 ± 8.58E-06 | 0.951 ± 0.010 | 0.981 ± 0.006 | 0.924 ± 0.024 | **310 ± 130** |
|  | 2.73E-06 ± 5.40E-06 | 0.983 ± 0.005 |  |  |  |
|  | 4.95E-05 ± 2.52E-06 | 0.930 ± 0.003 |  |  |  |
| **K102E + para-CFT** | 8.86E-05 ± 9.18E-06 | 0.853 ± 0.011 | 0.942 ± 0.008 | 0.785 ± 0.028 | **71 ± 16** |
|  | 1.72E-04 ± 2.11E-05 | 0.741 ± 0.026 |  |  |  |
|  | 1.28E-04 ± 1.40E-05 | 0.803 ± 0.016 |  |  |  |
| **W39A** | 1.43E-05 ± 3.82E-06 | 0.956 ± 0.004 | 0.990 ± 0.004 | 0.962 ± 0.016 | 690 ± 320 |
|  | 5.73E-06 ± 6.41E-06 | 0.979 ± 0.007 |  |  |  |
|  | 2.56E-05 ± 1.16E-05 | 0.955 ± 0.014 |  |  |  |
| **W39A + para-CFT** | 1.05E-04 ± 1.01E-05 | 0.848 ± 0.012 | 0.980 ± 0.006 | 0.922 ± 0.022 | **300 ± 110** |
|  | 1.70E-04 ± 2.66E-05 | 0.754 ± 0.030 |  |  |  |
|  | 4.05E-05 ± 2.19E-05 | 0.908 ± 0.025 |  |  |  |

Supplementary Figures

| | | $K_D$ / nM [1] |
|---|---|---|
| strA | 5' ATTGTAGTTTTTTTGAGTTGATTATGTTTTTAGTA 3' | |
| cstrA(16-26) | 3' TCAACTAATAC 5' | 370 ± 50 |
| cstrA(18-26) | 3' AACTAATAC 5' | 8500 ± 700 |
| cstrA(21-35) | 3' TAATACAAAATCAT 5' | 31.2 ± 0.8 |
| cstrA(26-35) | 3' CAAAATCAT 5' | 3900 ± 500 |

| | | $K_D$ / nM [1] |
|---|---|---|
| strB | 5' TTTTGTGAGATTACGGAACCTTTTTTTTTTTTTGT 3' | |
| cstrB(5-16) | 3' CACTCTAATGCC 5' | 20.1 ± 1.8 |
| cstrB(6-15) | 3' ACTCTAATGC 5' | 353 ± 7 |
| cstrB(8-16) | 3' TCTAATGCC 5' | 2120 ± 60 |

| | | $K_D$ / nM [1] |
|---|---|---|
| strC | 5' TTTTATCGGCTCTTCTCATTTTATTTTATTTTGTT 3' | |
| cstrC(7-16) | 3' GCCGAGAAGA 5' | 29 ± 7 |
| cstrC(8-15) | 3' CCGAGAAG 5' | 418 ± 16 |

**Figure S1: Sequences and $K_D$s of the DNAs used for qLILBID calibration.**
Sequences of three single-stranded DNAs (ssDNA), strA, strB, and strC, each consisting of 35 nucleotides (nt), displayed in the 5' to 3' orientation. Below each large strand, shorter ssDNA sequences (cstrA, cstrB, cstrC) are aligned in the 3' to 5' orientation to highlight the respective complementary binding regions to the larger strands within a distinct segment indicated by nucleotide position in parentheses: cstrA(16-26), cstrA(18-26), cstrA(21-35), and cstrA(26-35) bind to strA; cstrB(5-16), cstrB(6-15), and cstrB(8-16) bindto strB; and cstrC(7-16) and cstrC(8-15) bind to strC. Dissociation constant ($K_D$) values are listed on the right All $K_D$ values are taken from Young et al.[1]

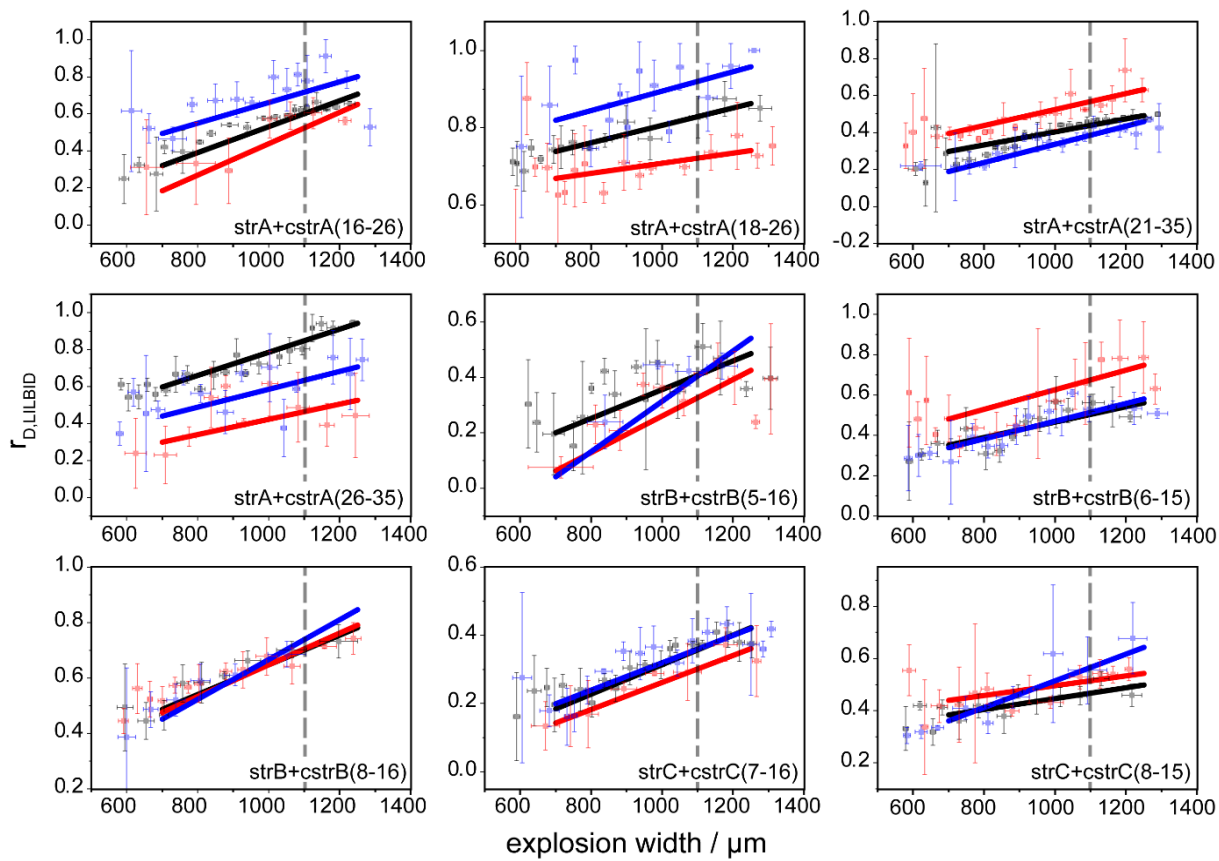

**Figure S2: Dissociation plots of dsDNAs used for qLILBID affinity calibration.**
Dissociation plots of strA, strB and strC and shorter strands complementary to different parts of the main strand (see **Figure S1**). The relative amount of monomer is plotted against the explosion width (plume size) of the droplet explosion and fitted linearly. Resulting fit parameters and percentage of dissociated complex at 1100 µm explosion width are given in **Supplementary Table 1**. For all samples, the measurements were repeated three times (shown in black, blue and red).

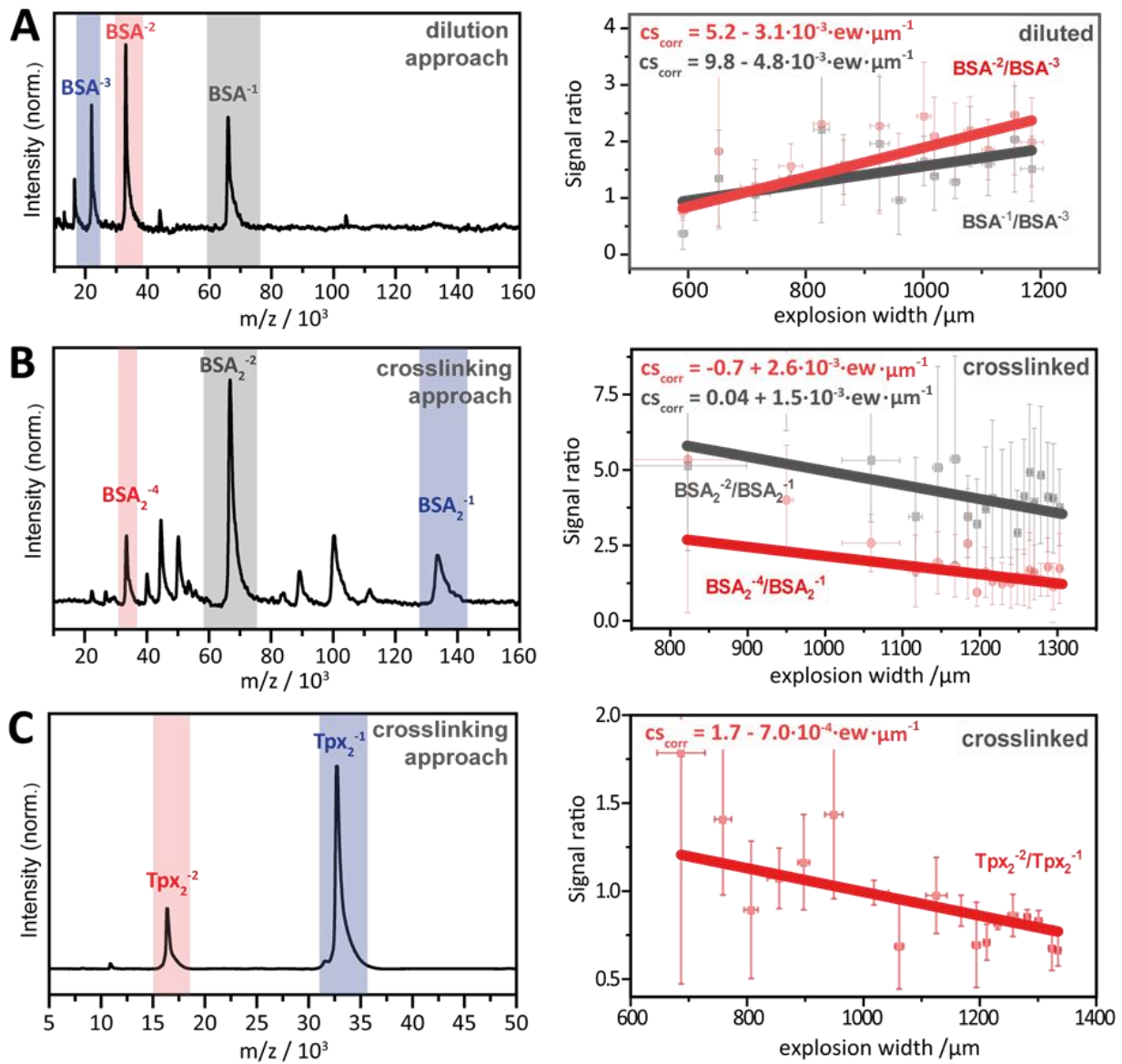

**Figure S3: Dilution and crosslinking approach with different protein homooligomers.**
*Left:* Spectra of either the crosslinked or diluted protein sample (Bovine Serum Albumin, BSA; Tryparedoxin, Tpx), where only one of the relevant species (either monomer or dimer) is visible. The relevant peak areas are highlighted. Blue is the reference peak, where no peak overlap in an equilibrium is expected. *Right:* The signal ratios gained from diluted or crosslinked spectra plotted against the explosion width. This is used to determine the charge state ratios as detailed in the main text. The fit yields a charge state correlation function ($cs_{corr}$), which is dependent on the explosion width (ew). **(A)** *Left:* Spectrum of diluted BSA (100 nM) with only monomeric species. *Right*: The signal ratio of doubly charged BSA (red) and singly charged BSA (grey) compared to the three times charged BSA (blue in spectrum) are plotted against the explosion width. **(B)** *Left:* Spectrum of the crosslinked BSA shows no monomeric species. The highlighted areas correspond to the relevant peak m/z. *Right*: The signal ratio of the four times charged BSA dimer (red) and two times charges dimer (grey) to the singly charged BSA dimer (blue in spectrum) are plotted against the explosion width. **(C)** *Left:* Spectrum of the crosslinked Tpx showing no monomeric species. The highlighted areas correspond to the relevant peak m/z. *Right*: Signal ratio gained from the diluted spectra to determine the monomeric and dimeric share in overlapping peaks. The signal ratio of the two times charged Tpx dimer (red) to the singly charged Tpx dimer (blue in spectrum) is plotted against the explosion width.

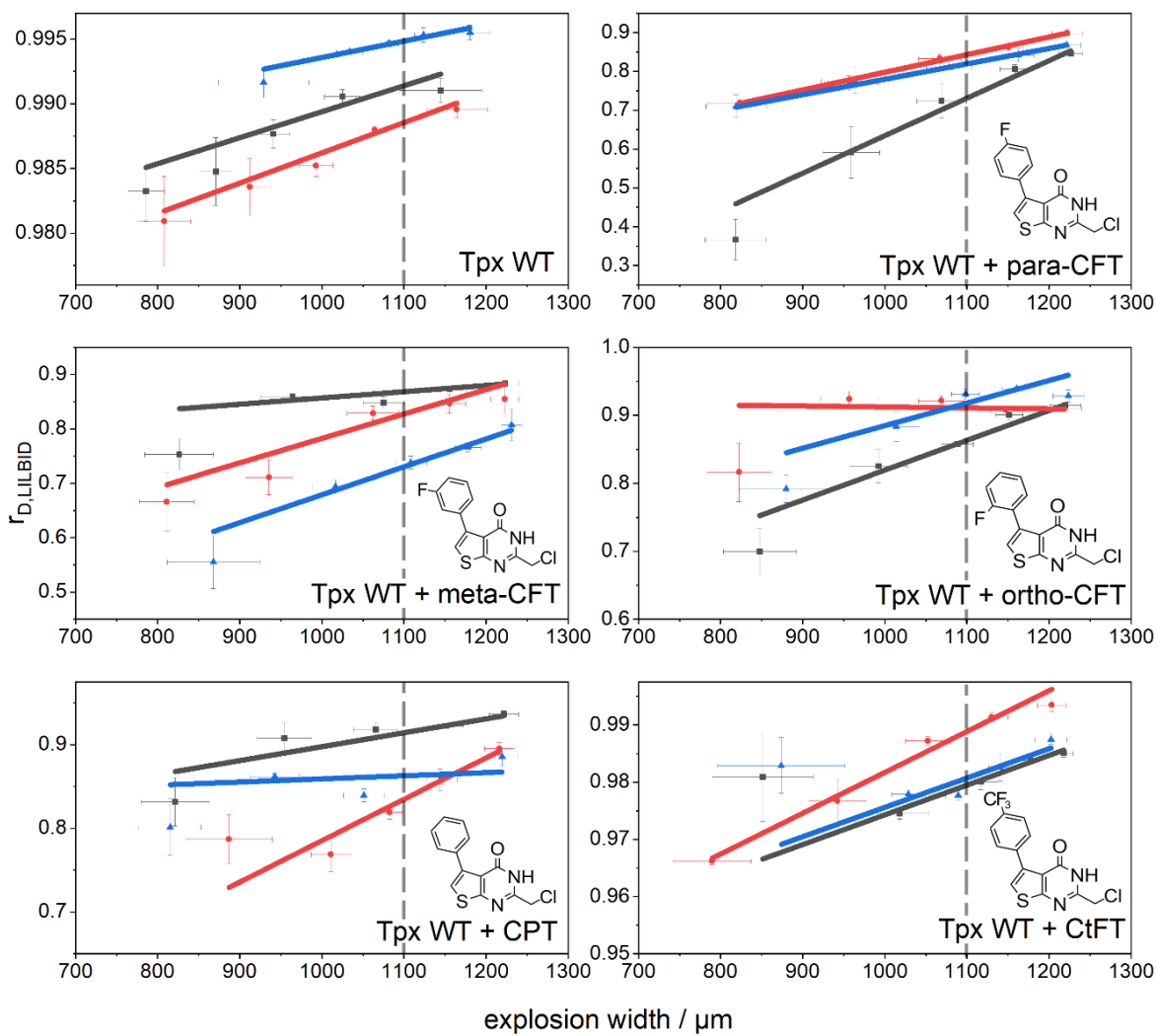

**Figure S4: Dissociation plots of Tpx WT with and without different covalently bound molecular glues.** Dissociation plots of Tpx WT alone and in complex with different covalent dimerizers, i.e. para-CFT, meta-CFT, ortho-CFT, CtFT and CPT. Depending on the molecular glue, each induced Tpx dimer has a different affinity.[2] The relative amount of monomer is plotted against the explosion width (plume size of the droplet explosion) and fitted linearly. Resulting fit parameters and percentage of dissociated complex at 1100 µm explosion width are given in **Supplementary Table 2**. For all samples, the measurements were repeated three times (each replicate shown in black, blue and red, respectively).

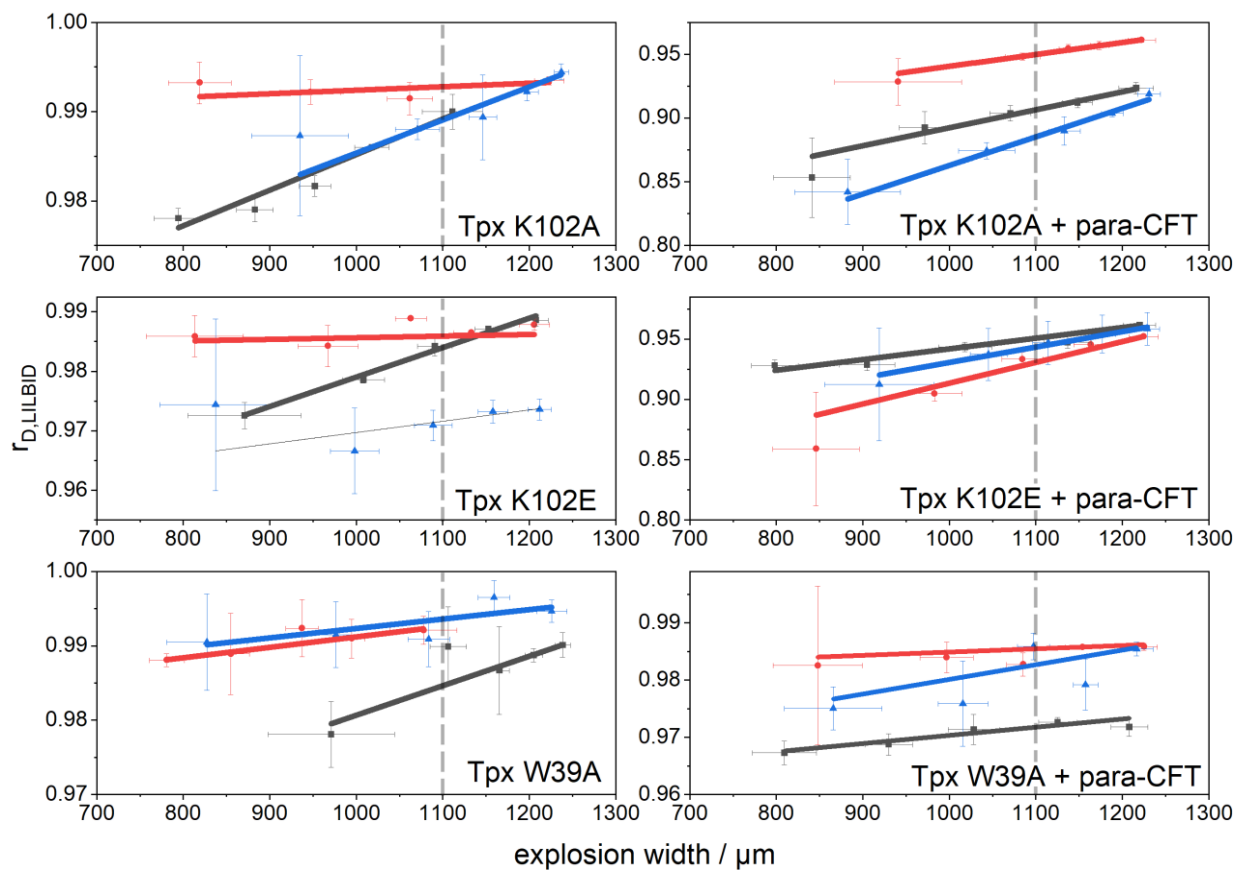

**Figure S5: Dissociation plots of Tpx K102 and W39 mutants with and without the covalently bound para-CFT inhibitor.**

Disssociation plots of Tpx point mutants K102E, K102A and W39A alone and in the presence of the molecular glue para-CFT. The mutants display significantly reduced dimer affinity compared to the WT. The relative amount of monomer is plotted against the explosion width (plume size of the droplet explosion) and fitted linearly. Resulting fit parameters and percentage of dissociated complex at 1100 µm explosion width are given in **Supplementary Tables 2 and 3**. For all samples, the measurements were repeated three times (each replicate shown in black, blue and red, respectively).

## Calculation of $r_{D,LILBID}$ from peak integrals

In BSA and Tpx spectra, we see charge state overlap. For BSA, the peaks at ca. 66 kDa and 33 kDa are peaks of overlapping species (Dimer$^{2-}$/Monomer$^{1-}$ and Dimer$^{4-}$/Monomer$^{2-}$ respectively) while for Tpx the peak at ca. 16 kDa is a peak of overlapping species (Dimer$^{2-}$/Monomer$^{1-}$). To distinguish between monomeric and dimeric contributions within the overlapping peaks, we applied charge state correlation functions to resolve their respective distributions. ($cs_{corr}$, **Figure S3**).

Crosslinking approach: (**Figure S3B, S3C**)

$$\int Dimer^{2-} = cs_{corr,1}(ew) \cdot \int Dimer^{1-} \tag{S1}$$

$$\int Monomer^{1-} = \int (Monomer^{1-} + Dimer^{2-}) - cs_{corr,1}(ew) \cdot \int Dimer^{1-} \tag{S2}$$

$$\int Dimer^{4-} = cs_{corr,2}(ew) \cdot \int Dimer^{1-} \tag{S3}$$

$$\int Monomer^{2-} = \int (Monomer^{2-} + Dimer^{4-}) - cs_{corr,2}(ew) \cdot \int Dimer^{1-} \tag{S4}$$

Dilution approach: (**Figure S3A**)

$$\int Monomer^{1-} = cs_{corr,1}(ew) \cdot \int Monomer^{3-} \tag{S5}$$

$$\int Dimer^{2-} = \int (Monomer^{1-} + Dimer^{2-}) - cs_{corr,1}(ew) \cdot \int Monomer^{3-} \tag{S6}$$

$$\int Monomer^{2-} = cs_{corr,2}(ew) \cdot \int Monomer^{3-} \tag{S7}$$

$$\int Dimer^{4-} = \int (Monomer^{2-} + Dimer^{4-}) - cs_{corr,2}(ew) \cdot \int Monomer^{3-} \tag{S8}$$

For all approaches and measurements:

$$r_{D,LILBID} = \frac{\sum_{cs} \int Monomer^{cs}}{\sum_{cs} \int Monomer^{cs} + \sum_{cs} \int Dimer^{cs}} \tag{S9}$$

**Derivation of the $K_D$ from $r_{D,solution}$ for heterodimers (equation 4a):**

$$r_{D,solution} = \frac{n(M_1)}{n(M_1 + D)} = \frac{n(M_2)}{n(M_2 + D)} \tag{3}$$

Using the same concentration for both binding partners ($c(M_1) = c(M_2) = c(M)$ and $c_{total}$ as the total concentration of each protein: $c_{total} = c(M) + c(D)$

$$r_{D,solution} = \frac{c(M)}{c_{total}} \tag{S10a}$$

$$\Rightarrow c(M) = r_{D,solution} \cdot c_{total} \tag{S10b}$$

Insert into the definition for $K_D$:

$$K_D = \frac{c(M_1) \cdot c(M_2)}{c(D)} = \frac{c(M)^2}{c(D)} = \frac{c(M)^2}{c_{total} - c(M)} \tag{S11}$$

$$K_D = \frac{(r_{D,solution} \cdot c_{total})^2}{c_{total} - r_{D,solution} \cdot c_{total}} = \frac{c_{total} \cdot r_{D,solution}^2}{1 - r_{D,solution}} \tag{S12}$$

Rearranging the equation enables calculation of expected r$_{D,solution}$ values for samples with known literature $K_D$ values for calibration (see Table S1):

$$r_{D,solution} = -\frac{K_D}{2\,c_{total}} + \sqrt{\left(\frac{K_D}{2\,c_{total}}\right)^2 + \frac{K_D}{c_{total}}} \tag{S12b}$$

**Derivation of the $K_D$ from r$_{D,solution}$ for homodimers (equation 4b):**

$$r_{D,solution} = \frac{n(M)}{n(M+D)} = \frac{c(M)}{c(M) + c(D)} \tag{S13}$$

With $c(D) = \frac{1}{2}(c_{total} - c(M))$:

$$r_{D,solution} = \frac{c(M)}{c(M) + \frac{1}{2}(c_{total} - c(M))} = \frac{2 \cdot c(M)}{c(M) + c_{total}} \tag{S14a}$$

$$c(M) = \frac{r_{D,solution} \cdot c_{total}}{2 - r_{D,solution}} \tag{S14b}$$

$$c(D) = \frac{1}{2}\left(c_{total} - \frac{r_{D,solution} \cdot c_{total}}{2 - r_{D,solution}}\right) = \frac{1}{2} c_{total} \cdot \left(1 - \frac{r_{D,solution}}{2 - r_{D,solution}}\right) \tag{S15a}$$

$$c(D) = \frac{1}{2} c_{total} \cdot \left(\frac{2 - r_{D,solution} - r_{D,solution}}{2 - r_{D,solution}}\right) = c_{total} \cdot \left(\frac{1 - r_{D,solution}}{2 - r_{D,solution}}\right) \tag{S15b}$$

Insert into the definition for $K_D$:

$$K_D = \frac{c(M) \cdot c(M)}{c(D)} = \frac{c(M)^2}{c(D)} = \frac{\left(\frac{r_{D,solution} \cdot c_{total}}{2 - r_{D,solution}}\right)^2}{c_{total} \cdot \left(\frac{1 - r_{D,solution}}{2 - r_{D,solution}}\right)} \tag{S16a}$$

$$K_D = \frac{c_0 \cdot \left(\frac{r_{D,solution}}{2 - r_{D,solution}}\right)^2}{\left(\frac{1 - r_{D,solution}}{2 - r_{D,solution}}\right)} = \frac{\frac{c_{total}}{2 - r_{D,solution}} \cdot r_{D,solution}^2}{1 - r_{D,solution}} \tag{S16b}$$

$$K_D = \frac{c_{total} \cdot r_{D,solution}^2}{r_{D,solution}^2 - 3 \cdot r_{D,solution} + 2} \tag{S16c}$$

.

**Materials and Methods**

**DNA preparation.** ssDNA sequences for calibration originate from Young et al.[1] and were purchased from Thermo Fisher Scientific as desalted, dry custom oligonucleotides (see Table 1 for sequences and $K_D$s). The set of ssDNAs comprised three non-self-complementary strands with lengths of 35 nucleobases, as well as nine shorter ssDNAs complementary to specific parts of the larger strands and differing in their length and C-G content. Each ssDNA was dissolved in a buffer containing 0.5 mM $MgHPO_4$ at pH 7.2 to reach a final concentration of 10 µM. To create dsDNAs with nM to low-µM affinities, equimolar amounts of long and short, complementary ssDNAs were annealed at 95°C for 10 minutes and gradually cooled to room temperature over a period of several hours.

**BSA preparation.** BSA (bovine serum albumin) was purchased from Sigma-Aldrich as a lyophilized powder and dissolved in 100 mM $NH_4CH_3COO$. To get solely dimeric BSA as reference, 20 µl of 30 µM BSA was crosslinked by incubating it with 2 mM 1-ethyl-3-[3-dimethylaminopropyl]-carbodiimid-hydrochlorid (EDC) in Tris-HCl buffer at pH 7 overnight and separating the dimer from unreacted BSA monomer and EDC by repeated dilution and filtering over an Amicon Ultra Centrifugal Filter unit (Sigma-Aldrich) with a cutoff of 100 kDa. Shortly before qLILBID-MS measurements, the crosslinked as well as the untreated BSA were desalted with Zeba™ Micro Spin desalting columns (Thermo Fisher Scientific) equilibrated with 200 mM $NH_4CH_3COO$ at pH 7.5.

**Tpx purification and derivatization.** *Trypanosoma brucei* Tryparedoxin (Tpx) was heterologously overexpressed as a proteolytically cleavable, $His_6$-tagged Trx-fusion protein in *E. coli* BL21GoldDE3 cells (Agilent) and purified via Ni-affinity (Qiagen) and size exclusion chromatography (SEC) (HiLoad16/600 Superdex75 pg column, Cytiva) as described previously.[2] For covalent modification, 100 µM Tpx were incubated with 4 mM tris(2-carboxyethyl)-phosphine (TCEP) and a 2-fold excess of dimerizer (4 mM in DMSO) or a 4-fold excess of crosslinker 1,8-bismaleimidodiethyleneglycol ($BM(PEG)_2$, 10 mM in DMSO) at 25 °C for 30 min, and purified via SEC (Superdex75 Increase 10/300 GL column (Cytiva) at 4 °C. All proteins were stored in Tpx buffer (25 mM NaPi, 150 mM NaCl, pH 7.5) at -20°C.

**ITC measurements.** Dissociation constants ($K_D$s) of Tpx constructs were determined via dilution ITC using MicroCal PEAQ-ITC instruments (Malvern Panalytical).[2] Tpx constructs in Tpx buffer were injected into sample cells with pure buffer, and the measured heat signals were analyzed using the MicroCal PEAQ-ITC Analysis software (Malvern Panalytical). Each titration was performed in a technical triplicate at room temperature.[2]

**LILBID laser dissociation plots.** LILBID-MS spectra of dsDNA and protein homodimers were acquired at room temperature using a custom-built mass spectrometer equipped with a LILBID ion source and Time-of-Flight (ToF) detector described in Morgner et al.[3] In brief, 5 µl of the aqueous sample are injected into a droplet generratror and droplets with concentrations between 0.1 to 30 µM are emitted through a glass capillary with a nozzle width of 50 µm. These droplets are then transferred into a vacuum environment of approximately $10^{-5}$ mbar and exposed to a 2.8 µm laser pulse (Innolas Spitlight 400, Continuum PowerLite 8000) lasting about 6 ns. This process causes a rapid expansion of the droplets, enabling ions from the sample to transition into the gas phase and be analyzed based on their mass-to-charge ratio (m/z). The resulting droplet plume is illuminated by a Minilite I laser (Continuum, San Jose, USA) 5 µs after the infrared (IR) laser pulse and captured by a DFK 23UP031 camera (Imaging Source, Bremen, Germany). Here, each droplet yielded a mass spectrum and an image of the corresponding plume. The size of the plume (explosion width, ew) was assessed using OpenCV with a self-written Python script. The IR laser energy was set to 32 mJ and the overlap between laser and droplet was varied to achieve droplet explosion widths from 600 to 1300 µm.